\documentstyle[12pt]{article}
\begin{document}
\begin{titlepage}
\begin{normalsize}
\flushright{UT-660 \\ November 1993} \\
\end{normalsize}
\vfil
\begin{LARGE}
\begin{center}
{
Explicit Relation of Quantum Hall Effect and Calogero-Sutherland Model
}
\end{center}
\end{LARGE}
\vfil
\begin{center}
\begin{Large}
{Hiroo Azuma and Satoshi Iso}
\vfil
\end{Large}
{Department of Physics, Faculty of Science, \\
University of Tokyo, \\
7-3-1 Hongo, Bunkyo-ku, Tokyo 113, Japan } \\
\vfil
Abstract
\end{center}
Explicit relation between Laughlin state of
the quantum Hall effect and one-dimensional(1D)
model with long-ranged interaction ($1/r^2$) is discussed.
By rewriting lowest Landau level wave functions in terms of 1D
representation, Laughlin state can be written as a deformation of the
ground state of Calogero-Sutherland model.
Corresponding to Laughlin state on different geometries, different
types of 1D $1/r^2$ interaction models are derived.
\end{titlepage}
\vfil
\newpage
\section{Introduction}
Recent studies of one-dimensional (1D) integrable models with
$1/r^2$ interaction \cite{calogero} \cite{sutherland}
show that the model has a close similarity
to the quantum Hall effect (QHE) \cite{laughlin}.
In both models, the ground state is given
by the Jastrow form and the excited states are constructed by
multiplying  polynomials to the ground state. Besides wave functions,
some properties common in both models  have been discussed.
It is  pointed out that the excited states of Haldane-Shastry model
(spin model with $1/r^2$ interaction) have fractional statistics as
the quasi-particle excitation in QHE \cite{haldane}.
Hierarchy extension of 1D model is also studied by
Kawakami \cite{kawakami}. He constructed a generalized 1D $1/r^2$
model with the same hierarchy as the QHE and showed that the matrix
classifying the excitation  is the same as the topological
order matrix introduced in the QHE.
 And the same algebraic structure ($W_{1+\infty}$ algebra)  has been studied
in both models to characterize their
 universal structure \cite{w} \cite{hikami}.  In spite of these similarities,
it has not yet
been clarified if there is any explicit relation between them.
\par
In this paper, we show that there is indeed an explicit relationship between
these two different models.
If two-dimensional electrons in strong magnetic field
 are constrained in the lowest Landau level, two of
four phase space degrees of freedom are frozen and effective
degrees are reduced.  Hence, wave functions in the
lowest Landau level can be represented in 1D form.
This makes it possible to relate 2D QHE with 1D $1/r^2$ model.
In this 1D representation,
Laughlin states of FQHE on different
geometries are shown to correspond to different 1D models with $1/r^2$
interaction.
Laughlin state  on disk
is shown to be rewritten as a one-parameter deformation of the ground state
of Calogero model (1D integrable model with  $1/r^2$ interaction
in harmonic potential).  The deformation parameter is the magnetic
field $B$. The Laughlin wave function on cylinder is rewritten
as a deformation of the ground state of Sutherland model (periodic 1D model
with
$1/\sin(r)^2$
interaction without harmonic potential).
\par
By these correspondences,
it is shown that both models have many common properties.
Excitations in 1D $1/r^2$ model corresponding to quasi-holes (particles)
in QHE must have fractional charge and statistics.
Moreover, the same algebraic structure ($W_{1+\infty}$ algebra)
will characterize their universal structure.

\section{Kinematics}
For a planar electron in a magnetic field normal to plane Hamiltonian is given
by
\begin{equation}
H_0=\sum_{i=1,2} {(\Pi_i)^2 \over 2m}, \ \ \ \Pi_i=p_i-A_i, \ \ \ i=1,2.
\end{equation}
Here we set $c=\hbar=e=1$ and assume that the constant magnetic field
is in the negative $z$ direction. By defining an annihilation operator
\begin{equation}
a=(\Pi_x-i \Pi_y)/\sqrt{2B}, \ \ \ [a,a^{\dagger}]=1,
\end{equation}
$H_0$ is written as
$H_0=\omega_c (a^{\dagger} a+1/2),$  $\omega_c=B/m$.
Heisenberg  equation of motion
\begin{equation}
\dot{\pi_x}=i[H_0,\pi_x]=-\omega_c \pi_y, \ \
\dot{\pi_y}=i[H_0,\pi_y]=\omega_c \pi_x
\end{equation}
show that $(\pi_x,\pi_y)$ rotate with frequency $\omega_c$
and therefore
represent cyclotron motion in a magnetic field.
Spectrum of $H_0$ is  quantized as Landau levels. States in the lowest
Landau level (LLL) satisfy the LLL condition $a\phi_0=0$ and
$a^{\dagger}$ creates states in higher Landau levels.
The guiding center coordinates
of the cyclotron motion are defined by
\begin{equation}
X=x-{\Pi_y \over B}, \ \ \ Y=y+{\Pi_x \over B},\ \ \
[X,Y]={i \over B}.
\end{equation}
They commute with $a$ and $a^{\dagger}$ and therefore with $H_0$.
These coordinates describe degeneracy in each Landau level.
These four variables $(\Pi_x, \Pi_y, X, Y)$ are  more convenient
phase space variables than $(p_x, p_y, x, y)$
in a constant magnetic field.
 \\

In a strong magnetic field, all electrons are confined in
LLL whose one-particle wave functions satisfy the
LLL constraint $a \phi=0$.  If we constrain the Hilbert space onto
the LLL, two of four phase space variables, $(\Pi_x, \Pi_y)$ are
frozen and remaining degrees of freedom are
the guiding center coordinates
 $(X,Y)$.  Therefore, effective degrees of freedom in the LLL
are reduced to the half of the total degrees of freedom
 and the two-dimensional
$(X,Y)$-coordinate space can be seen as a phase space of 1D system
\cite{iso}.
That is, when $X$ is diagonalized $X|s \rangle =s|s \rangle$,  Y is interpreted
as its dual momentum $Y=p_s/B$.
 $|s \rangle$, eigenstate of $X$, is uniquely determined with the
LLL condition and forms a complete  basis in the LLL;
\begin{equation}
\int |s \rangle \langle s| ds=1.
\end{equation}
Since any LLL wave function can be written in terms of
$|s \rangle$, wave functions and Hamiltonian in LLL can be interpreted as
those of 1D system whose coordinate is $s$.
In the following, we study s-representation ( or 1D
representation) of LLL wave functions.
\\
\section{1D representation of disk Laughlin state}
First let's consider 1D representation of one-particle LLL wave function
in symmetric gauge; ${\bf A}=(By/2, -Bx/2)$.
In this gauge, the annihilation operator $a$ is given by
$a=-i(\partial_z +\bar{z}/2)$ where $z=\sqrt{B/2}(x+iy)$ and
the LLL wave functions are written as
$\langle z \bar{z}|\Psi \rangle =\Psi(\bar{z}) e^{-|z|^2/2}$.
Normalized eigenfunction $|s \rangle$ of the guiding center coordinate
$X=i \partial_y/B+x/2$ in the LLL
is given by
\begin{equation}
{1 \over \sqrt{2\pi}}
\langle z \bar{z}|s \rangle =\left( {B \over \pi}  \right)^{1/4}
e^{-Bs^2/2} e^{\sqrt{2B} s\bar{z}-\bar{z}^2/2}e^{-|z|^2/2}.
\label{s-wf1}
\end{equation}
{}From equation (\ref{s-wf1}), 1D representation of LLL wave function
$\langle z \bar{z}| \Psi \rangle =\Psi (\bar{z}) e^{-|z|^2/2}$ is given by
\begin{eqnarray}
\langle s|\Psi \rangle
&=& \int \langle s|z \bar{z} \rangle \langle z \bar{z}|\Psi \rangle
{d^2 z \over \pi} \nonumber \\
&=& {1 \over \sqrt{2\pi}}
\left( {B \over \pi}  \right) ^{1/4} \int
e^{-Bs^2/2} e^{\sqrt{2B} s z -z^2/2}e^{-|z|^2} \Psi(\bar{z})
{d^2 z \over \pi} \nonumber \\
&=&{1 \over \sqrt{2\pi}}
\left( {B \over \pi}  \right) ^{1/4}
e^{-Bs^2/2} e^{-{1 \over 4B} \left( {\partial \over \partial s} \right) ^2}
\int e^{\sqrt{2B} s z -|z|^2} \Psi(\bar{z})
{d^2 z\over \pi} \nonumber \\
&=& {1 \over \sqrt{2\pi}}
\left( {B \over \pi}  \right)^{1/4}
e^{-Bs^2/2} e^{-{1 \over 4B} \left( {\partial \over \partial s} \right)^2}
\Psi(\sqrt{2B}s).
\label{s-rep1}
\end{eqnarray}
In the last equality, we used the coherent state identity
\begin{equation}
\int e^{\alpha z-|z|^2} \Psi(\bar{z}) {d^2 z \over \pi} =\Psi(\alpha).
\end{equation}
Using 1D representation of LLL wave functions, dynamics in LLL can be
interpreted as dynamics of 1D system.
\par
Now we generalize  (\ref{s-rep1}) to many-particles
case. N-particles wave functions in the LLL are generally written as
\begin{equation}
\langle z_1 \bar{z_1} \cdot \cdot \cdot z_N \bar{z_N}|\Psi \rangle
=\Psi (\bar{z_1} \cdot \cdot \cdot \bar{z_N}) e^{-\sum_i |z_i|^2/2}.
\label{wf}
\end{equation}
 1D representation of the wave function (\ref{wf}) is
\begin{equation}
\langle s_1 \cdot \cdot \cdot s_N|\Psi \rangle =
\left( {B \over 4} \right) ^{N/4}
e^{-B \sum_i s_i^2/2}
e^{-{1 \over 4B} \sum_i \left( {\partial \over \partial s_i} \right)^2}
\Psi(\sqrt{2B}s_1,  \cdot \cdot \cdot ,\sqrt{2B} s_N).
\end{equation}
For the Laughlin state
$ \Psi_m(\bar{z_1} \cdot \cdot \cdot \bar{z_N}) =
\prod_{i<j} (\bar{z_i}-\bar{z_j})^m $,
1D representation becomes
\begin{equation}
\langle s_1 \cdot \cdot \cdot s_N|\Psi \rangle =
e^{-B \sum_i s_i^2/2}
e^{-{1 \over 4B} \sum_i \left( {\partial \over \partial s_i} \right)^2}
\prod_{i<j} (s_i-s_j)^m.
\label{laugh1}
\end{equation}
Here we neglected a constant normalization factor for simplicity.
For $m=1$,  as is expected, this is the Slater determinant of the lowest
N eigenstates of a harmonic oscillator.
\par
The 1D representation of Laughlin state (\ref{laugh1}) has an interesting
property.
Momentum representation of (\ref{laugh1}) is given by the same form
\begin{equation}
\langle t_1 \cdot \cdot \cdot t_N|\Psi \rangle =
e^{-B \sum_i t_i^2/2}
e^{-{1 \over 4B} \sum_i \left( {\partial \over \partial t_i} \right)^2}
\prod_{i<j} (t_i-t_j)^m
\label{laugh1t}
\end{equation}
where $\langle s|t \rangle=e^{iBts}/\sqrt{2\pi }$ (momentum $p$ is set by
$p_i=Bt_i$.)
This duality is due to rotational invariance of the Laughlin state on disk
(circular
droplet).
In momentum representation, guiding center coordinate $Y$ is diagonalized
by $Y|t \rangle =t|t \rangle$ and therefore 1D representation (\ref{laugh1})
($X$ is taken as a 1D coordinate) and  (\ref{laugh1t}) ($Y$ is taken as a
1D coordinate) must have the same form.
Here we comment on rotational invariance. Rotation generator for guiding center
coordinates is given by
\begin{equation}
R=\sum_i \left\{ {B \over 2}(X_i^2+Y_i^2)-{1 \over 2}  \right\}, \ \
[R,X_i]=-iY_i, \ \  [R,Y_i]=iX_i.
\end{equation}
It is easy to prove that
\begin{eqnarray}
&&R e^{-B \sum_i s_i^2/2}
e^{-{1 \over 4B} \sum_i \left( {\partial \over \partial s_i} \right)^2}
f(s_1, \cdot \cdot \cdot ,s_N) \nonumber \\
&& = e^{-B \sum_i s_i^2/2}
e^{-{1 \over 4B} \sum_i \left( {\partial \over \partial s_i} \right)^2}
(\sum_i s_i \left( {\partial \over \partial s_i} \right) )
f(s_1, \cdot \cdot \cdot ,s_N).
\end{eqnarray}
Therefore  LLL state
$e^{-B \sum_i s_i^2/2}
e^{-{1 \over 4B} \sum_i \left( {\partial \over \partial ,s_i} \right)^2}
f(s_1, \cdot \cdot \cdot s_N)$ is rotational invariant if $f(s_1, \cdot \cdot
\cdot s_N)$
is a homogeneous function of $s_i$. \par
Rewriting  (\ref{laugh1}) by dimensionless parameter $\tilde{s}=\sqrt{B}s$,
no dimensionful parameter as magnetic field $B$ exists.
Long distance behaviour ($\sqrt{B}|s_i-s_j|=|\tilde{s_i}-\tilde{s_j}| \gg 1$)
of the wave function eq.(\ref{laugh1}) is described by the wave function
($B \rightarrow \infty$ limit of (\ref{laugh1}))
\begin{equation}
\langle s_1 \cdot \cdot \cdot s_N|\Psi \rangle =
e^{-B \sum_i s_i^2/2}
\prod_{i<j} (s_i-s_j)^m.
\end{equation}
It is an exact form for $m=1$ state (\ref{laugh1}).
This is the well-known groundstate wave function of 1D integrable
model with $1/r^2$ interaction in harmonic potential (Calogero model);
\begin{equation}
H=\sum_i {p_i^2 \over 2} +\sum_{i<j} {m^2-m \over (s_i-s_j)^2 }
+\sum_i {B^2 s_i^2 \over 2}.
\end{equation}
Short distance behaviour ($ |\tilde{s_i}-\tilde{s_j}| \ll 1$), on the other
hand, is described by the ground state of Calogero model in
$t$-space \cite{brink}. \par
Excited states also correspond between QHE and Calogero model in 1D.
Quasi-holes in QHE are constructed by multiplying $\prod_i
(\bar{z_i}-\bar{z_0})$
on the Laughlin state. Then, in 1D representation, this excited state is
constructed
by multiplying $\prod_i (\sqrt{2B} s_i-\bar{z_0})$ on the Calogero ground
state.
Since the quasi-hole has 1/m fractional statistics and fractional charge
independent of the magnetic field or the shape of the droplet, corresponding
excited states in 1D $1/r^2$ model also have the same fractional statistics and
fractional charge.

\section{1D representation of cylinder Laughlin state}
Next let's consider 1D representation of Laughlin state on cylinder.
Here we use Landau gauge for convenience ${\bf A}=(By, 0)$.
In this gauge, LLL wave functions are written as $\Psi(\bar{z}) e^{-By^2/2}$.
We impose a periodic boundary condition for $x$ with period $L_x$.
Then an anti-holomorphic part of a LLL wave function can be written as
a linear combination of $\exp[2\pi i n   (x-iy)/L_x]=\omega^n$
where $\omega \equiv \exp[2\pi i   (x-iy)/ L_x] .$
  Filling factor $\nu=1$ state is given by
\begin{equation}
\prod_{i<j} (\omega_i-\omega_j) e^{-B \sum_i y_i^2/2}.
\end{equation}
Since $\prod_{i<j} (\omega_i-\omega_j)$ is a Slater determinant
of $(1,\omega, ...,\omega^{N-1})$, there are two boundaries at $y=0$
and $y=2\pi (N-1)/BL_x$.
Laughlin state with filling factor $\nu=1/m$ can be constructed as
\begin{equation}
\prod_{i<j} (\omega_i-\omega_j)^m  e^{-B \sum_i y_i^2/2}.
\label{laugh2}
\end{equation}
Note that its short distance behaviour is the same as the disk Laughlin
state.  Filled region is expanded by $m$-times and
 boundaries are located at $y=0$ and $y=2\pi m(N-1)/BL_x$.
\par
Now let's consider 1D representation of eq.(\ref{laugh2}).
In Landau gauge, eigenstate of $X=x+i \partial_y /B$ in LLL is
given by
\begin{equation}
\langle z \bar{z}|s \rangle ={1 \over \sqrt{2\pi}}
\left( {B \over \pi}  \right)^{1/4}
e^{-Bs^2/2} e^{\sqrt{2B} s\bar{z}-\bar{z}^2} e^{-By^2/2}.
\end{equation}
Since we identify $s$ and $s+L_x$, we must sum all $s$ mod
$L_x$;
\begin{equation}
|s \rangle \rightarrow |s \rangle_{per.} \equiv \sum_n |s+nL_x \rangle.
\end{equation}
Then $\langle z \bar{z}|s \rangle_{per.}$ is shown to be
under a shift $x \rightarrow x+L_x$ and can be written as a
linear combination of $\omega^n$.
 1D representation of LLL wave function
$
\langle z \bar{z}|\Psi \rangle=\Psi(\bar{z}) e^{-By^2/2}
$
is
\begin{eqnarray}
\langle s|\Psi \rangle
&=& \int \langle s|z \bar{z} \rangle \langle z \bar{z}|\Psi \rangle
{d^2 z \over \pi} \nonumber \\
&=& {1 \over \sqrt{2\pi}}
\left( {B \over \pi}  \right) ^{1/4} \int
e^{-Bs^2/2} e^{\sqrt{2B} s z -{z^2 \over 2}}e^{-|z|^2}
e^{\bar{z}^2/2} \Psi(\bar{z})
{d^2 z \over \pi} \nonumber \\
&=&  {1 \over \sqrt{2\pi}}
\left( {B \over \pi}  \right)^{1/4}
e^{-Bs^2/2} e^{-{1 \over 4B} \left( {\partial \over \partial s} \right)^2}
e^{Bs^2} \Psi(\sqrt{2B}s)  \nonumber \\
&=& {1 \over \sqrt{2\pi}}
\left( {B \over \pi}  \right)^{1/4}
e^{-Bs^2/2} e^{-{1 \over 4B} \left( {\partial \over \partial s} \right)^2}
e^{Bs^2} 2^{s \partial_s} \Psi(\sqrt{B / 2}s)
\nonumber \\
&=& {1 \over \sqrt{2\pi}}
\left( {B \over \pi}  \right)^{1/4} g_0 \
e^{-{1 \over 2B} \left( {\partial \over \partial s} \right)^2}
\Psi(\sqrt{B /2}s) ,
\label{cylinder1}
\end{eqnarray}
where
$g_0=\sum_{n=0}^{\infty} (-1/4)^n (2n)!/(n!)^2$.
By the replacement $\bar{z} \rightarrow \sqrt{B/2} s$, $\omega$ becomes
$e^{2\pi i s/L_x}$ and invariant under shift $s \rightarrow s+L_x$.
Therefore, in the following, we use $|s \rangle$ instead of
$|s \rangle_{per.}$  for notational simplicity.
Extending it to many-particle case,
 1D representation of Laughlin state on cylinder
is given by
\begin{equation}
\langle s_1 \cdot \cdot \cdot s_N|\Psi \rangle =
e^{-{1 \over 2B} \sum_i \left( {\partial \over \partial s_i} \right)^2}
\prod_{i<j} (e^{i2\pi s_i/L_x} -e^{i2\pi s_j/L_x} )^m.
\end{equation}
In $B \rightarrow \infty$ limit, this wave function reduces to the
ground state of Sutherland model (periodic 1D model with
  $1/\sin^2(\pi(s_i-s_j)/L_x)$ interaction) \cite{yoshioka};
\begin{equation}
\prod_{i<j} (e^{i2\pi s_i/L_x} -e^{i2\pi s_j/L_x} )^m.
\end{equation}
For $m=1$, $\sum_i (\partial_i)^2$ becomes a constant and this
is exact.
In strong magnetic field limit, the width of cylinder Laughlin state
$\delta y=2\pi (N-1)m/BL_x$ becomes infinitesimal.
In disk case, Laughlin state is reduced to 1D system in harmonic potential.
But Laughlin state on cylinder is reduced to 1D system without external
potential. This is due to the difference in shape of the droplet on
two-dimensional phase space.
\par
Now let's study the momentum representation, or $Y$-diagonalized
representation of (\ref{cylinder1}). Set $|t \rangle$ by
$\langle s|t \rangle=e^{iBts}/\sqrt{2\pi}$ as before.
Then Fourier-transformation of (\ref{cylinder1}) gives
\begin{equation}
\langle t|\Psi \rangle=
{1 \over \sqrt{2\pi B}}
\left( {B \over \pi}  \right)^{1/4}
e^{-Bt^2/2} e^{-{1 \over 4B} \left( {\partial \over \partial t} \right)^2}
e^{-Bt^2}\Psi(-i\sqrt{2B}t).
\label{cylinder2}
\end{equation}
To derive it, we used the second form of eq. (\ref{cylinder1}).
Rewriting (\ref{cylinder2}), it becomes
\begin{equation}
\langle t|\Psi \rangle=
{1 \over \sqrt{2\pi B}}
\left( {B \over \pi}  \right)^{1/4}  g_0 \
e^{-{1 \over 2B} \left( {\partial \over \partial t} \right)^2}
e^{-Bt^2/2}\Psi(-i\sqrt{B/2}t).
\end{equation}
By the replacement $\bar{z} \rightarrow -i\sqrt{B/2} t$,
$\omega =\exp[2\pi i  / L_x (x-iy)] $ becomes
$\exp(2\pi t/L_x)$.
Therefore, Laughlin state on cylinder  (\ref{laugh2}) has the following
$t$-representaion;
\begin{eqnarray}
\langle t_1 \cdot \cdot \cdot t_N|\Psi \rangle  &=&
e^{-{1 \over 2B} \sum_i \left( {\partial \over \partial t_i} \right)^2}
e^{-B \sum_i t_i^2/2}
\prod_{i<j} (e^{2\pi t_i/L_x} -e^{2\pi t_j/L_x} )^m \nonumber \\
& \propto &
e^{-{1 \over 2B} \sum_i \left( {\partial \over \partial t_i} \right)^2}
e^{-B \sum_i (t_i-t_0)^2/2}
\prod_{i<j} (\sinh \pi (t_i-t_j)/L_x)^m, \nonumber \\
&&
\end{eqnarray}
where $t_0 \equiv m (N-1)\pi /B L_x$.
In large $B$ limit, this becomes
\begin{equation}
 e^{-B \sum_i (t_i-t_0)^2/2}
\prod_{i<j} (\sinh \pi (t_i-t_j)/L_x)^m.
\end{equation}
Note that  $t_0$ is the
center of the two boundaries ( at $Y=0$ and $Y=2\pi m(N-1)/B L_x$ )
of the cylinder Laughlin droplet.

\section{Conclusion}
In this letter, we presented an explicit relation between Laughlin state
and one-dimensional integrable model with $1/r^2$ interaction.
In one-dimensional representation of lowest Landau level wave
functions, Laughlin state can be written as a one-parameter
deformation of the ground state of 1D model $1/r^2$ model.
Different types of 1D models are derived ($1/r^2$, $1/\sin ^2 r$
and $1/\sinh ^2 r$) corresponding to Laughlin state on different
geometries. The deformation parameter is magnetic field $B$.
\par
Finally we list some  topics which we will discuss in separate papers
\cite{iso2}. \\
(1) Laughlin state on torus and its 1D model \\
(2) $W_{1+\infty}$ algebra in QHE and 1D $1/r^2$ model \\
(3) extension to hierarchy and  $SU(N)$ generalization of 1D model\\
(4) Tomonaga-Luttinger liquid behaviour of Laughlin state and relation with
edge state\\
(5) $X$-$Y$ duality in two-dimensions and duality of long-distance and short
distance physics in one-dimension.
\par
We gratefully acknowlegde helpful discussion with B. Sakita on
QHE and with K. Hikami on Calogero-Sutherland  model.
S. Iso is supported by Grant-in Aid for Scientific Research from
the Ministry of Education, Science and Culture in Japan.

\end{document}